\pdfoutput=1
\documentclass[12pt,aps,superscriptaddress]{revtex4-2}

\usepackage{amsmath,amssymb}
\usepackage{bm}
\usepackage{graphicx}
\usepackage[colorlinks=true,citecolor=blue,urlcolor=blue,linkcolor=blue,bookmarksopen]{hyperref}

\newcommand\<{\langle}
\renewcommand\>{\rangle}
\renewcommand\Im{\mathop{\mathrm{Im}}}
\newcommand\x{\mathbf{x}}

\newcommand\+{\dagger}
\newcommand\p{\bm{p}}

\newcommand\U{{\mathcal{U}}}

\begin{document}

\title{Unnuclear physics}

\author{Hans-Werner Hammer}
\affiliation{Technische Universit\"{a}t Darmstadt, Department of Physics, Institut f\"{u}r Kernphysik, 64289 Darmstadt, Germany}
\affiliation{ExtreMe Matter Institute EMMI and Helmholtz Forschungsakademie Hessen für FAIR (HFHF), GSI Helmholtzzentrum f\"{u}r Schwerionenforschung GmbH,
64291 Darmstadt, Germany}

\author{Dam Thanh Son}
\affiliation{Kadanoff Center for Theoretical Physics, Enrico Fermi Institute, and James Franck Institute, University of Chicago, Chicago, IL 60637, USA}

\begin{abstract}
We investigate a nonrelativistic version of Georgi's ``unparticle
physics.''  We define the unnucleus as a field in a nonrelativistic
conformal field theory.  Such a field is characterized by a mass and a conformal
dimension.  We then consider the formal problem of scatterings to a
final state consisting of a particle and an unnucleus and show that
the differential cross section, as a function of the recoil energy
received by the particle, has a power-law singularity near the maximal
recoil energy, where the power is determined by the conformal
dimension of the unnucleus.  We argue that unlike the relativistic
unparticle, which remains a hypothetical object, the unnucleus is
realized, to a good approximation, in nuclear reactions involving
emission of a few neutrons, when the energy of the final-state
neutrons in their center-of-mass frame lies in the range between about
$0.1$~MeV and 5~MeV.  Combining this observation with the known
universal properties of fermions at unitarity in a harmonic trap, we
predict a power-law behavior of an inclusive cross section in this
kinematic regime.  We compare our predictions with previous
effective field theory and model calculations of the
$^6$He$(p,p\alpha)2n$, $^3$H$(\pi^-,\gamma)3n$, and
$^3$H$(\mu^-,\nu_\mu)3n$ reactions and find excellent agreement.
\end{abstract}

\maketitle

\section{Introduction}

In a 2007 paper~\cite{Georgi:2007ek} Howard Georgi proposed the
concept of an ``unparticle,'' which gave rise to considerable
activity in theoretical particle physics.  Georgi's idea is that,
beyond the Standard Model, there is a hidden sector consisting not of
particles, but of fields belonging to a conformal field theory.  In
general, correlation functions of fields in conformal field theory do
not have poles, but only cuts, so the ``unparticles'' that correspond
to these fields, if they exist, would leave collider signals distinct from
those of the normal particles.  Despite intensive search, so far the
unparticle has failed to turn up at the
LHC~\cite{Khachatryan:2014rra,Khachatryan:2015bbl,Sirunyan:2017onm}.

In this paper, we consider a nonrelativistic analog of the unparticle,
which we call the ``unnucleus.''  Formally, the unnucleus corresponds
to a field in a nonrelativistic conformal field
theory~\cite{Nishida:2007pj}.  In contrast to the relativistic
unparticle, which is characterized solely by its conformal dimension,
the nonrelativistic counterpart is characterized by two
parameters---its mass $M$ and dimension $\Delta$.  We use the term
``unnucleus'' because, as we will argue later, this object appears in
a certain regime in nuclear reactions involving several neutrons in
the final state.  Thus, in contrast to the unparticle, the unnuclei
already exist in nature, although only as an approximation.

Our result can be summarized as follows.  Consider a nuclear reaction
with a few final-state neutrons, beside one other product which we
call $B$, for example
\begin{equation}
  A_1 + A_2 \to B + \underbrace{n + n + \cdots}_{N~\text{neutrons}}
\end{equation}
The number of final-state neutrons $N$ can be $2, 3, 4,\ldots$.  We
register only the energy of $B$, but not of the neutrons, measuring
the inclusive
differential cross section as the function of the energy $E$
of $B$ and its direction $\Omega$, $d^2\sigma/dEd\Omega$.  In the
center-of-mass frame the rate does not depend on the direction of $B$,
so what is measured is $d\sigma/dE$.
The energy spectrum of $B$ is continuous and has a
cutoff at some maximal value $E_0$.  We predict that
\begin{equation}
  \frac{d\sigma}{dE} \sim (E_0-E)^\nu
\end{equation}
with some exponent $\nu$ that depends on the number of final-state
neutrons, in the regime where
\begin{equation}
  \frac{\hbar^2}{m a^2} \ll \left(1+ \frac{M_B}{N m}\right) (E_0-E)
  \ll \frac{\hbar^2} {m r_0^2} \,,
\end{equation}
where $m$ is the mass of the neutron, $M_B$ the mass of the nucleus
$B$, while $a$ and $r_0$ are the neutron-neutron scattering length and
effective range.  If $M_B$ is not too large compared to $Nm$, this
means $E_0-E$ is between 0.1~MeV and a few MeV.

The exponent $\nu$ is predicted to be
\begin{equation}
  \nu = \begin{cases}
     -1/2 & N = 2,\\
     1.77 & N =3, \\
     2.5-2.6 & N=4.
  \end{cases}
\end{equation}
In general, $\nu$ is equal to the ground state energy of a system of $N$
fermions at unitarity in a harmonic trap with unit frequency, minus
$\frac52$.

The structure of this paper is as follows.  In
Sec.~\ref{sec:unnucleus} we introduce the notion of an unnucleus and
review the properties of the unnucleus propagator as followed from
nonrelativistic conformal field theory.  In Sec.~\ref{sec:rate} we
compute the rate of processes involving an unnucleus in the final
state.  We argue that multi-neutron final states can approximate
unnuclei in Sec.~\ref{sec:multi-neutron}.  In
Sec.~\ref{sec:comparison} we compare our prediction
for multi-neutron spectra with previous model
calculations for several nuclear reactions.
Finally, Sec.~\ref{sec:conclusion} contains concluding remarks.

\section{The unnucleus}
\label{sec:unnucleus}

We will start our discussion at a rather formal level and transition
to real nuclear processes later.

The unnucleus is a nonrelativistic field with mass $M$ and dimension
$\Delta$.  There is a unitarity bound on $\Delta$: $\Delta\ge\frac32$,
where the lower bound corresponds to a free field.  (In our convention,
the dimensions of momentum  and energy are 1 and 2, respectively.)
According to the
general formalism, the two-point function of a primary operator $\U$
in nonrelativistic conformal field theory is completely fixed (up to an overall
factor), so the propagator of an unnucleus is~\cite{Nishida:2007pj}
\begin{equation}\label{U-prop}
  G_\U(t, \x) =
  -i \< T \U(t,\x) \U^\+(0,\mathbf{0}) \> =
  C \, \frac{\theta(t)}{(it)^{\Delta}}
  \exp\left ( \frac{iM x^2}{2t}\right),
\end{equation}
where $C$ is a normalization factor.  For
$\Delta=\frac32$ (the dimension of a free field), the unnucleus
becomes a nucleus (a nonrelativistic particle).

One example of the unnucleus is a collection of noninteracting particles,
\begin{equation}\label{composite}
  \U = \psi_1 \psi_2 \cdots \psi_N .
\end{equation}
Assuming the masses of all all fields $\phi_i$ are equal, the mass and
the dimension of this operator are
\begin{equation}
  M = N m_\psi, \qquad \Delta = \frac32 N .
\end{equation}
The propagator of $\U$ is then the $N$th power of the propagator of a
single particle.

For diagrammatic calculation we need the unnucleus propagator in
momentum space.  Taking the Fourier transform of Eq.~(\ref{U-prop}) we
get
\begin{equation}
  G_\U(\omega,\p) = - C  \left( \frac{2\pi}M\right)^{3/2}
  \Gamma\left( \frac52-\Delta\right)
  \left( \frac{p^2}{2M}-\omega \right)^{\Delta-\frac52} .
\end{equation}
In Fourier space the imaginary part of the propagator
of an unnucleus is
\begin{equation}\label{ImG}
  \Im G_\U(\omega, \p) \sim
  \begin{cases}
  \delta\Bigl( \omega - \frac{p^2}{2M}\Bigr),
  & \Delta = \frac32 \,, \\
  \left( \omega - \frac{p^2}{2M} \right)^{\Delta-\frac52}
  \theta \Bigl( \omega - \frac{p^2}{2M}\Bigr),
  & \Delta > \frac32 \,.
  \end{cases}
\end{equation}
Only for free fields ($\Delta=\frac32$) the propagator has a pole,
otherwise it has a cut.  For the composite operator~(\ref{composite}),
$\Im G_\U$ is proportional to the final-state phase space available
when an initial state carrying energy $E$ and momentum $\p$ becomes
$N$ final particles.  Similar to the relativistic case, an unnucleus
of dimension $\Delta$ can be thought of as $N=\frac23\Delta$ (which
is, in general, a fractional number) particles.  The imaginary part
of the unnucleus propagator can be interpreted as the phase space
volume of a fractional number of particles.

\section{Rate of processes involving an unnucleus}
\label{sec:rate}

\begin{figure}[ht]
 \centering
  \includegraphics[width=0.35 \textwidth]{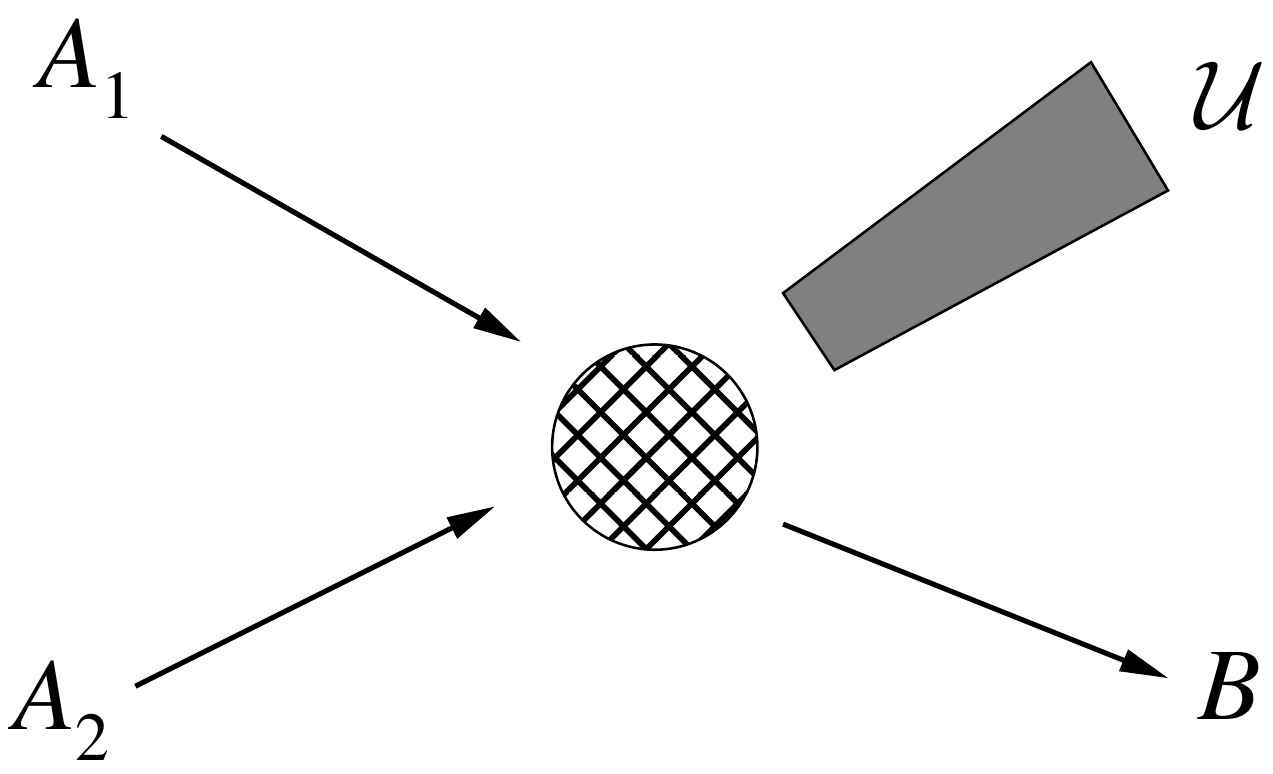}
 \caption{A nuclear reaction with an unnucleus  ${\mathcal U}$
 (represented by the shaded region) in the final state.}
  \label{fig:unnuclear-reaction}
\end{figure}

To illustrate the physical consequences of the existence of an
unnucleus, consider the following reaction (see
Fig.~\ref{fig:unnuclear-reaction})
\begin{equation}\label{process}
  A_1 + A_2 \to B + \mathcal U , 
\end{equation}
where $A_1$ and $A_2$ are some initial particles, $B$ is a particle
and $\mathcal U$ is the unnucleus.  For simplicity, we assume all
particles involved in the reaction are nonrelativistic, though our
main conclusion requires that only $\U$ is.  We work in the
center-of-mass frame.  The total kinetic energy available to final
products is
\begin{equation}
  E_{\text{kin}} = (M_{A_1} + M_{A_2} - M_B - M_{\mathcal U}) c^2
    + \frac{p_{A_1}^2}{M_{A_1}} + \frac{p_{A_2}^2}{M_{A_2}} \,.
\end{equation}
Unless $\mathcal U$ is a particle, the energy spectrum of $B$ is
continuous.  Let $E$ and $\p$ be the energy of the particle $B$,
$E=p^2/2m_B$.  We are interested in the differential cross section
$d\sigma/dE$.  We can think about a term in the effective Lagrangian
\begin{equation}\label{L-g}
  \mathcal L_{\text{int}} = g\,\U^\+ B^\+ A_1 A_2 + \text{h.c.}
\end{equation}
where $g$ is some coupling constant.  The differential cross section
can be computed to be
\begin{equation}\label{factorization}
  \frac{d\sigma}{dE}
  \sim 
  |\mathcal M|^2
  \sqrt{E}
  \Im G_\U (E_{\text{kin}} {-} E, \p). 
\end{equation}
For the Lagrangian~(\ref{L-g}) $\mathcal M=g$, but in principle
$\mathcal M$ can contain dependence on the momenta of the incoming and
outgoing particles.  The statement of Eq.~(\ref{factorization}) is
that the cross section can be factorized into two parts, one (encoded
by $\mathcal M$) corresponding to the primary process $A_1+A_2\to
B+\U$, the other (encoded by $\Im G_\U$) corresponding to the final-state
interaction between the constituents of $\U$.
Such a factorization requires that the energy scale of the primary
scattering process is much larger than that of the interaction between
the neutrons and 
is the essence of
the Watson-Migdal approach to final-state
interaction~\cite{Watson:1952ji,Migdal:1955}.

According to Eq.~(\ref{ImG}),
\begin{equation}
  \Im G_\U (E_{\text{kin}} {-} E, \p)
  \sim \left( E_{\text{kin}} - E - \frac{p^2}{2M_\U}\right)^{\Delta-\frac52}
  = \left[ E_{\text{kin}}
  - \left(1 + \frac{M_B}{M_\U}\right)E \right]^{\Delta-\frac52} .
\end{equation}
Denote the maximal value of the recoil energy received by the particle $B$ as
\begin{equation}
  E_0 = \left(1 + \frac{M_B}{M_\U}\right)^{-1} E_{\text{kin}} .
\end{equation}
In the regime $E_0-E\ll E_0$, 
ignoring the energy dependence of all other factors, we can write
\begin{equation}\label{scaling}
  \frac{d\sigma}{dE} \sim (E_0 - E)^{\Delta-\frac52} .
\end{equation}
Thus, a characteristic feature of processes involving an unnucleus is
the power-law dependence of the differential cross section on the
recoil energy near the end point.

\section{Multi-neutron final states as unnuclei}
\label{sec:multi-neutron}

So far the search for relativistic unparticles has been
unsuccessful~\cite{Khachatryan:2014rra,Khachatryan:2015bbl,Sirunyan:2017onm}.
In nuclear physics, however, there are natural approximate unnuclei
due to the fortuitous occurrence of fine tuning in several nuclear
systems.  In particular, neutrons have a large $s$-wave scattering
length: $a \approx -19$~fm, compared to the effective range $r_0\approx
2.8$~fm.  A system of neutrons can be considered as an unnucleus if
the relative momentum between any two neutrons in the system is
between $\hbar/a$ and $\hbar/r_0$. If this is the case,
they are described by a well
known nonrelativistic conformal field theory---the theory of fermions
at unitarity.

\begin{figure}[ht]
 \centering
  \includegraphics[width=0.35 \textwidth]{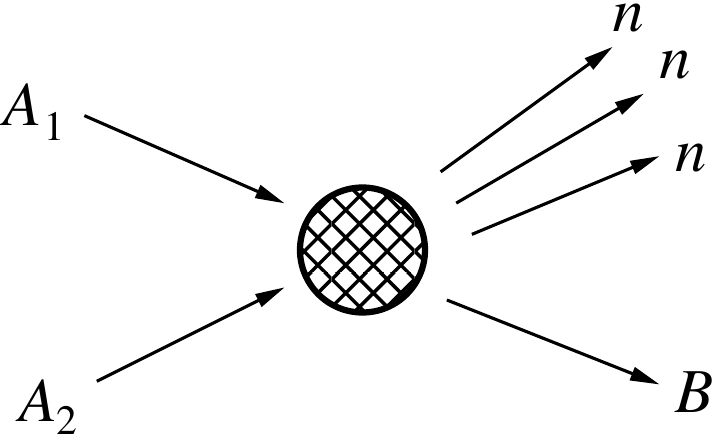}
 \caption{A nuclear reaction with three neutrons in the final state.}
  \label{fig:reaction}
\end{figure}

Thus, the real-world realizations of the reaction pictured in
Fig.~\ref{fig:unnuclear-reaction} are reactions with a few neutrons in
the final state. A typical reaction with three final-state
neutrons is schematically depicted in Fig.~\ref{fig:reaction}.  The
differential cross section $d\sigma/dE$ considered above is now an
inclusive cross section, where the momenta of the neutrons are left
unmeasured.  Reactions of this type are abundant in nuclear physics.
Some examples are
\begin{align}
  ^3\text{H} + {^3}\text H &\to {^4}\text{He} + 2n\,,\\
^7\text{Li} + {^7}\text{Li} &\to {^{11}}\text{C} + 3n\,,\\
  ^4\text{He} + {^8}\text{He} &\to {^8}\text{Be} + 4n\,.
\end{align}
The final-state neutrons can be considered as forming an unnucleus
when the kinetic energy of the system of neutrons in its
center-of-mass frame (neutron kinetic energy)
is between $\varepsilon_0=\hbar^2/ma^2\sim
0.1$~MeV and $\hbar^2/mr_0^2\sim 5$~MeV.
Only in this kinematic regime, our
prediction~(\ref{scaling}) for $d\sigma/dE$ applies.  Physically, in
this regime the neutrons travel together and keep interacting with
each other until the distance between them becomes larger than $a$.
If the total kinetic energy of the final scattering products
$E_{\text{kin}}$ is much larger than $\hbar^2/mr_0^2$, then the
power-law behavior of the differential cross section~(\ref{scaling})
is expected in a region near, but not too close to, the maximal recoil
energy.

According to the general formalism~\cite{Nishida:2007pj} the dimension
of an operator is equal to the energy of the corresponding state in
the harmonic potential with unit oscillator frequency.  This leads to
an nontrivial connection between the few-body physics of fermions at
unitarity and the physics of nuclear reactions.  Namely, the spectrum
of fermions at unitarity in a harmonic trap determines the behavior of
the processes involving emission of neutrons in a certain kinematic
regime.

For emission of two neutrons, the ground state of two unitary fermions
in a harmonic trap (with opposite spins) is known exactly, and
corresponds to the ``dimer'' operator of conformal dimension
$\Delta=2$.  The differential cross section thus grows toward the
endpoint
\begin{equation}
  \frac{d\sigma}{d E} \sim \frac1{\sqrt{E_0-E}} \,.
\end{equation}
This growth stops very close to the end point when the
neutron kinetic energy
is of order $\varepsilon_0$, after which, the two
neutrons become effectively noninteracting, and the unnucleus now
becomes an operator in free field theory $n_\downarrow n_\uparrow$
with dimension $\Delta=3$, and the differential cross section decreases
as $\sqrt{E_0-E}$.  This non-monotonic behavior of $d\sigma/dE$ is
well known~\cite{Migdal:1955}.  In fact, the whole behavior of the
differential cross section in the crossover region can be read off
from the propagator of the dimer field in effective field theory,
\begin{equation}
  G_d (\omega, 0) \sim \frac1{\frac1a + i\sqrt{m\omega}} \Rightarrow
  \Im G_d(\omega,0) \sim \frac{\sqrt\omega}{\varepsilon_0 + \omega}\,,
  \label{eq:dimer}
\end{equation}
so
\begin{equation}
  \frac{d\sigma}{d E} \sim \frac{\sqrt{E_0-E}}
  {E_0-E+ \left(1+\frac{M_B}{M_\U}\right)^{-1}\varepsilon_0}\,.
\end{equation}
which reaches a maximum at $E_0-E=(1+M_B/M_\U)^{-1}\varepsilon_0$.

For the problem of three final-state neutrons, we know that the ground
state of three fermions at unitarity in a harmonic trap corresponds is
a state with $S=\tfrac12$, $L=1$ and energy $\Delta\approx 4.27272$
in units of the trap frequency~\cite{Tan:2004,Werner:2006zz}.  Thus
the differential cross section behaves as
\begin{equation}\label{3n-scaling}
  \frac{d\sigma}{d E} \sim (E_0-E)^{1.77272} .
\end{equation}
The first excited state of three in the trap is a $S=\tfrac12$, $L=0$
state with $\Delta\approx4.66622$, corresponding to a contribution
$(E_0-E)^{2.1662}$.  This is suppressed compared to the contribution
from the ground state, but, due to the relatively small difference
between the exponents, may need to be taken into account to
describe real data.

At very small $E_0-E$ there is a crossover from Eq.~(\ref{3n-scaling})
to the free-neutron behavior, controlled by the dimension of the
operator $nn\nabla n$ in free field theory (with one derivative
because of the Pauli exclusion principle): $(E_0-E)^3$.  This behavior
can also be obtained by multiplying the three-particle phase space
$(E_0-E)^2$ and a suppression factor $E_0-E$ coming from the fermionic
statistics of the neutrons.

For four final-state neutrons, different approaches have given the ground-state energy of four trapped unitary fermions between $5.0$ and
$5.1$ oscillator frequencies~\cite{Chang:2007zzd,vonStecher:2007zz,Alhassid:2007cda,vonStecher:2009mu,Rotureau:2010uz,Endres:2011er,Rotureau:2013exe}, which means
\begin{equation}
  \frac{d\sigma}{d E} \sim (E_0-E)^{\alpha}, \quad \alpha\approx 2.5-2.6 .
\end{equation}
The excited state of the four-fermion system has $\Delta\approx6.6$~\cite{Rotureau:2013exe} and thus is much more separated
from the ground state compared to the three-fermion case.
The behavior crosses over to the free-particle behavior
$(E_0-E)^{5.5}$ at very low $E_0-E$.

We will not consider larger numbers of final-state neutrons, except to
point out that differential cross section will fall off with larger
and larger exponent as $E\to E_0$ with increasing number of
neutrons.

\section{Comparison with multi-neutron spectra}
\label{sec:comparison}

Ideally, one should compare our predictions with experimental measurements.
But since at present there are no sufficiently precise experimental
spectra in the endpoint region to identify a multi-neutron unnucleus, we
compare our predictions
to realistic theoretical calculations. For convenience, we consider
the neutron energy distribution in their center of mass instead of the
energy distribution of the recoil particle. This makes it possible
to consider reactions with more than one particle besides the neutrons
in the final state and also makes the relevant
energy scales more transparent. We expect that a comparison to precise
experimental low-energy two- and four-neutron spectra will
become possible in the
near future~\cite{nn_scat_len_ribf_prop2018,nnnn_ribf_prop2014}.

\begin{figure}[ht]
 \centering
  \includegraphics[width=0.9 \textwidth]{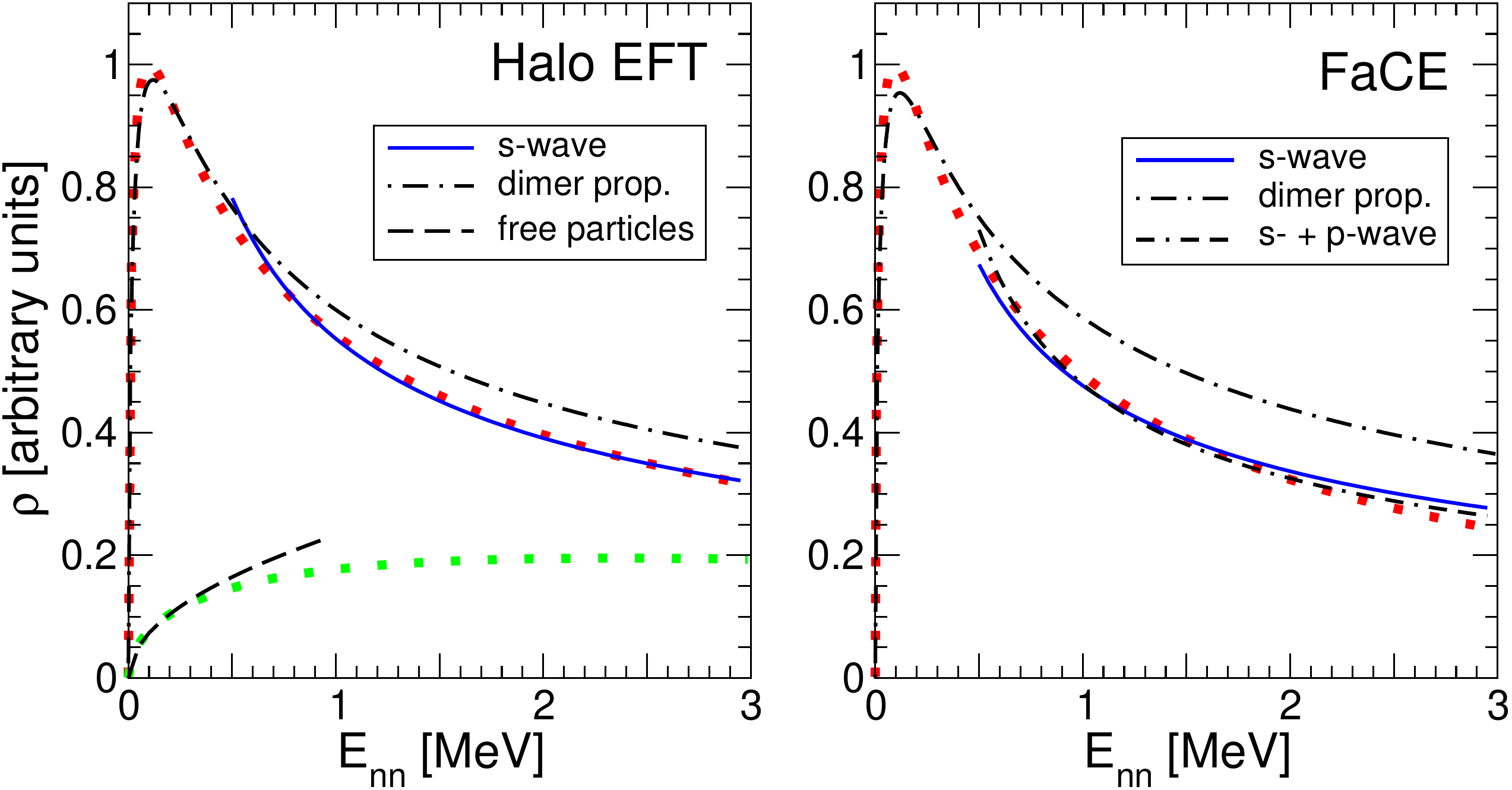}
 \caption{Center-of-mass energy spectrum of two neutrons in the reaction
 $^6$He$(p,p\alpha)2n$ at high energies.  Halo EFT calculations from G\"obel
 et al.~\cite{Gobel:2021pvw} with/without final state interaction of
 the neutrons are given by the upper red/lower green dotted lines
 in the left panel. FaCE calculations with final state interaction
 are given by the dotted line in the right panel. Different fits are
 explained in the legend and in the main text.}
  \label{fig:6He-Enn}
\end{figure}

In Ref.~\cite{Gobel:2021pvw}, a  novel method to measure the neutron-neutron
scattering length using the $^6$He$(p,p\alpha)2n$ in inverse kinematics at
high energies was proposed. It uses the final state interaction of
the neutrons after the sudden knockout of the $\alpha$ particle
in $^6$He. The authors showed that the scattering length can be extracted
from the spectrum of the
neutrons at very low relative energies. Here we use the two-neutron spectra
calculated in Ref.~\cite{Gobel:2021pvw} to search for the two-neutron
unnucleus. Once the data from the experiment \cite{nn_scat_len_ribf_prop2018}
are available, the analysis can be repeated using the measured spectrum.
In Fig.~\ref{fig:6He-Enn}, we analyze the calculations of
Ref.~\cite{Gobel:2021pvw}  with respect to signatures of the
two-neutron unparticle.
In that paper, calculations within two effective three-body approaches for the
wave function of the initial $^6$He nucleus are carried out:
(i) a leading order Halo effective field theory (Halo EFT) calculation which includes
$nn$ $s$-wave interactions, $n\alpha$ $p$-wave interactions,
and a short-range $nn\alpha$ three-body force \cite{Ji:2014wta}
(left panel) and
(ii) a model calculation using the three-body code FaCE
\cite{Thompson:2004dc} which
has $n\alpha$ interactions in the $s$-, $p$- and $d$-wave
and a longer ranged three-body force (right panel).
The two-neutron distribution from Halo EFT (upper red dotted line)
is well described by the unnucleus behavior, $1/\sqrt{E}$, above 0.5 MeV
as indicated by the solid line. In fact, even the full energy distribution
up to 3 MeV can reasonably well be described by the dimer propagator,
Eq.~(\ref{eq:dimer}) (dash-dotted line), by just fitting the prefactor
to the data below 0.5 MeV. If the propagator is fit to the whole
energy range, a better description at higher energies can be achieved
at the expense of a somewhat worse description of the peak.
The deviations are due to the initial $^6$He
wave function, which also enters into the description of the reaction.
This can be seen by the lower green dotted curve which gives the
energy distribution without the $nn$ final state interaction.
This distribution is well described by the free-neutron behavior
$\sqrt{E}$ up to about 0.5 MeV. At this energy scale, it seems that
structure effects from the $^6$He wave function become important
and the neutron distribution starts to differ from the free case.
This is consistent with the intrinsic scale generated by the two-neutron
separation energy of $^6$He, which is of order 1 MeV.
A similar behavior is observed in the FaCE
calculation in the right panel (dotted curve). However, in this case
the description of the FaCE calculation for energies beyond
0.5 MeV can be improved by also including the $p$-wave contribution 
which falls of as $1/E^{3/2}$ (dash-dash-dotted line). We expect this
to be due to
the more complicated structure of the $^6$He wave function in
FaCE, which also generates $p$-wave neutron pairs in the reaction.

\begin{figure}[ht]
 \centerline{
  \includegraphics[width=0.48 \textwidth]{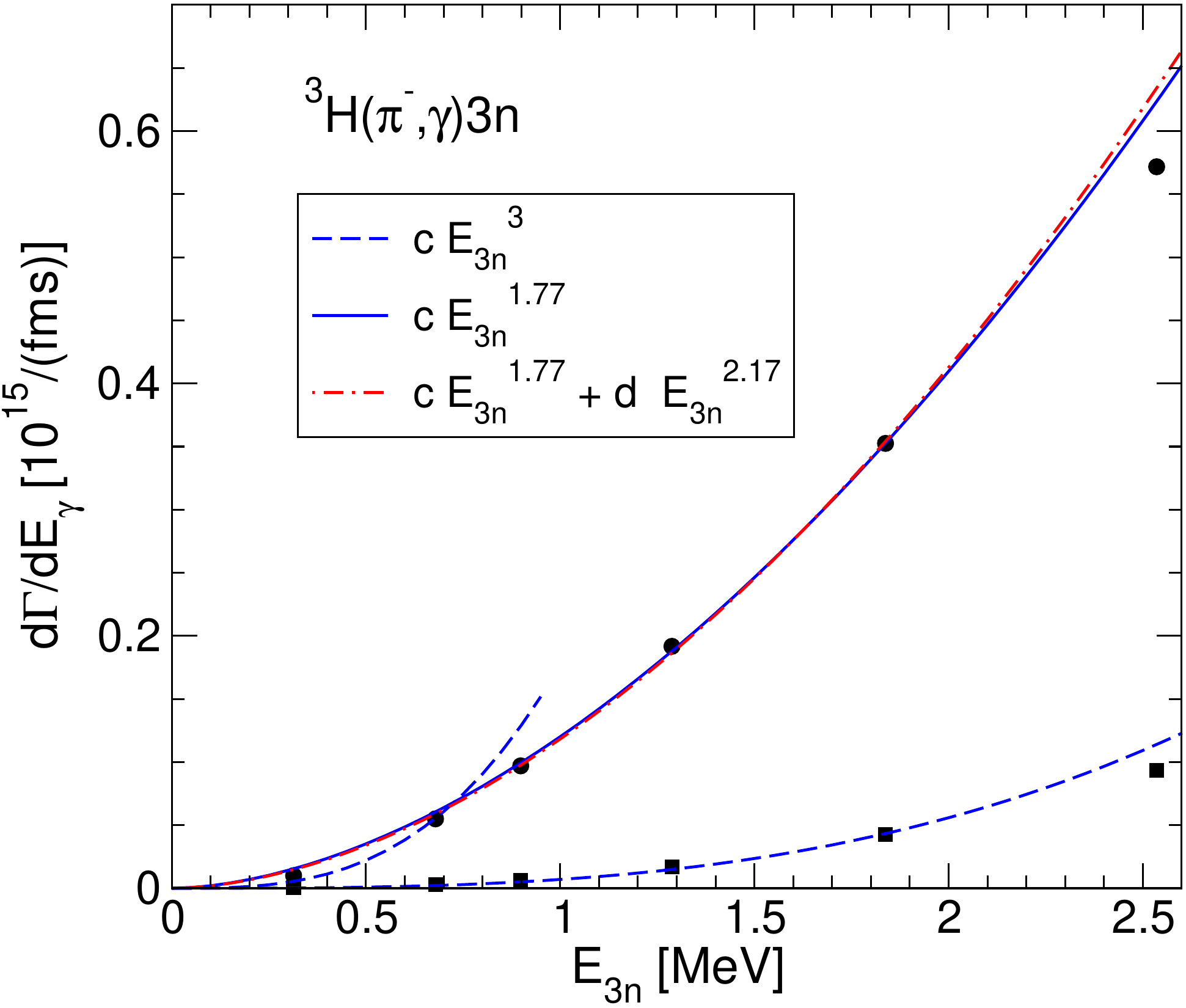}\quad
   \includegraphics[width=0.48 \textwidth]{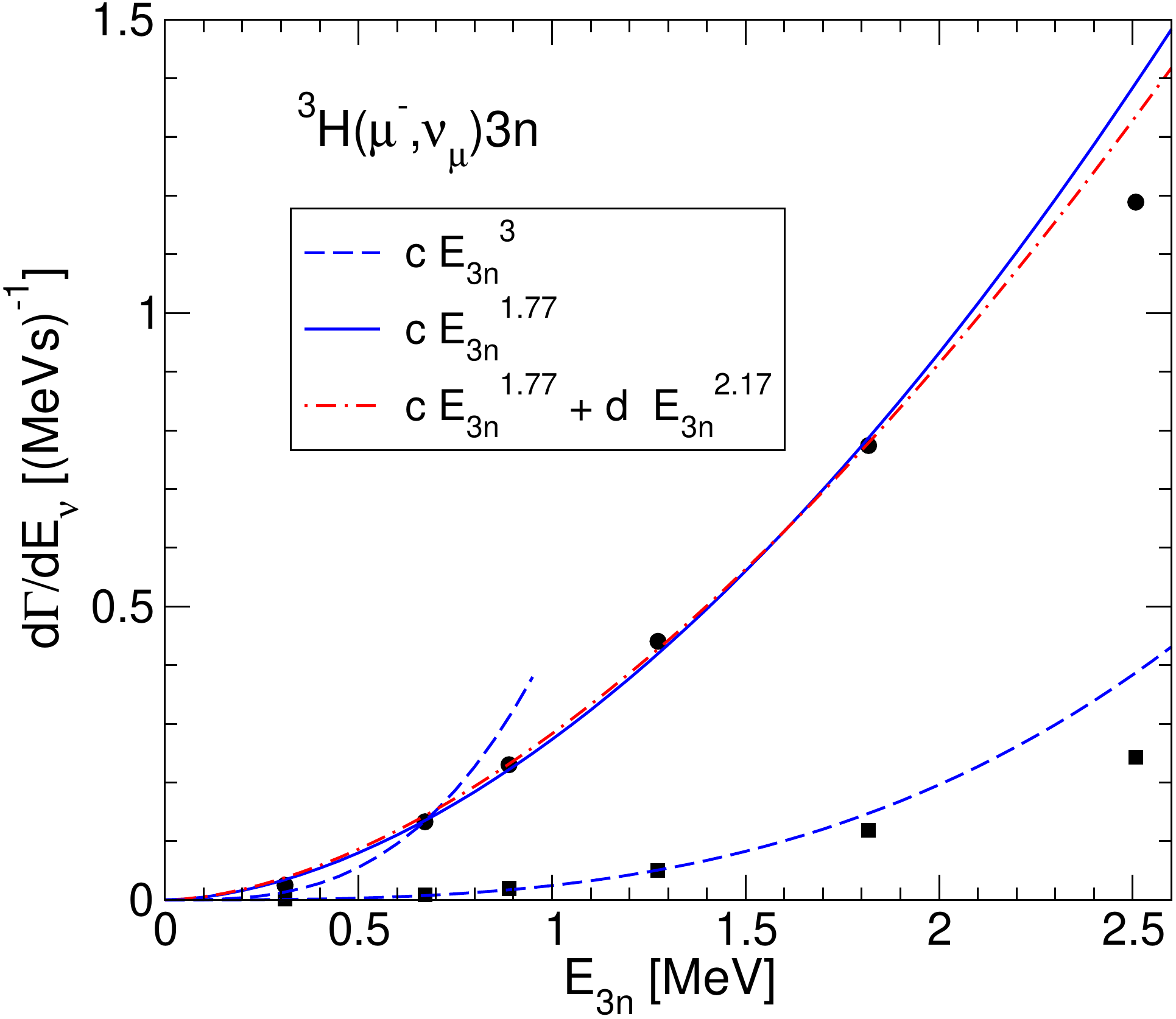}}
 \caption{Center-of-mass energy spectrum of three neutrons in the reaction
 $^3$H$(\pi^-,\gamma)3n$ (left panel) and $^3$H$(\mu^-,\nu_\mu)3n$
 (right panel). The circles/squares give the full/plane wave
 calculations by Golak et al.~\cite{Golak:2018jje,Golak:2016zcw}.
 Different fits are explained in the legend and in the main text.}
  \label{fig:pic-Ennn}
\end{figure}

Next we turn to the case of a
three-neutron final state. A precise photon spectrum near the
kinematical endpoint for radiative capture of stopped pions on
tritium was measured by Miller at al.~\cite{Miller:1980pn}.
While unnucleus behavior is consistent with the spectrum of Miller et
al., we cannot unambiguously extract
the power behavior from these data. Therefore, we turn to the theoretical
calculation of Golak et al.~\cite{Golak:2018jje}. They have
carried out a realistic model calculation of the   capture rate
for the reaction $^3$H$(\pi^-,\gamma)3n$ using the AV18 two-nucleon
potential and a Urbana IX three-body force. Their results are shown
in the left panel of Fig.~\ref{fig:pic-Ennn} for the full calculation (circles)
and the plane wave impulse approximation (squares). We have 
converted the calculated photon spectra to three-neutron spectra 
for convenience. As expected, the free neutron behavior,
$E^3$ (dashed line),
can describe the full calculation (circles)
only at the lowest energies. However,
the  plane wave impulse approximation (squares) can be described up to about
2.5 MeV.
The full calculation including final state interaction
displays clear unnucleus behavior, $E^{1.77}$ (solid line) for
energies also up to about 2.5 MeV, where it starts to deviate from
the prediction.  This is somewhat smaller than the value 5~MeV expected
from the scattering length.  We suspect that this is due to
the wave function of the triton, which
has finite extent,
making the reaction a less than ideal ``point source''
of the neutrons and causing the factorization formula~(\ref{factorization}) to
break down earlier than expected.
The description cannot be significantly improved by
including the next state which behaves as $E^{2.17}$ (dash-dotted line).
Analogous behavior is exhibited by the theoretical spectra
for the reaction  $^3$H$(\mu^-,\nu_\mu)3n$ calculated by Golak
et al.~\cite{Golak:2016zcw} using the same interaction model
(see right panel of Fig.~\ref{fig:pic-Ennn}). In this reaction,
the energy scale of the primary scattering process is slightly smaller
such that the corrections to factorization are larger.

A four-neutron spectrum was recently measured by Kisamori et al.
in the reaction $^4$He$(^8$He,$^8$Be$)4n$ \cite{Kisamori:2016jie},
but the number of events is too low to extract evidence of unnucleus
behavior. It may, however, be possible to extract such behavior from
the spectra of a new experiment using the reaction $^8$He$(p,p\alpha)4n$,
which are currently being analyzed \cite{nnnn_ribf_prop2014}.

\section{Conclusion}
\label{sec:conclusion}

We have suggested that nuclear processes involving a few neutrons in
the final state may be well described, in a certain kinematic regime,
as the production of an unnucleus, defined as an object corresponding
to a field in a nonrelativistic conformal field theory.  Using this
observation, we predict power-law behaviors of the differential cross
section in a certain range of the neutron kinetic energy, or equivalently, of
the recoil energy of the particle that emits the neutrons,
with the value of the exponent determined by the
universal physics of fermions at unitarity.

The power-law behavior breaks down when the relative momentum between
the neutrons is less than $\hbar/a$, crossing over to the regime of
free neutrons.  The transition between the two regimes is well known
in the case of two-neutron final state.  For final states containing
more than two neutrons, this crossover can be, in principle, studied
within the effective field theory approach.

The problem can be formalized as the calculation of the imaginary part
of the two-point Green's function of an operator $\U$ in the
nonrelativistic conformal field theory of unitary fermions,
deformed by a relevant deformation corresponding to a finite
scattering length $a$.  The theory therefore flows from an ultraviolet
fixed point of fermions at unitarity to an infrared free-fermion fixed
point.  We expect that in such a theory
\begin{equation}
  \Im G_\U(\omega, \mathbf{0})
  = \theta(\omega) \omega^{\Delta-\frac52}
  F_\U\Bigl( \frac\omega{\varepsilon_0}\Bigr),
\end{equation}
where $F_\U(\omega/\varepsilon_0)$ are universal functions, one for
each primary operator $\U$, which are expected to have the following
asymptotic behavior
\begin{equation}
  F_\U(x) \to \begin{cases}
   c_1 & x \gg 1 ,\\
   c_2 x^{\Delta_{\text{free}}-\Delta} & x \ll 1 ,
  \end{cases}
\end{equation}
where $c_1$ and $c_2$ are constant, and $\Delta_{\text{free}}$ is the
dimension of the operator that $\U$ becomes in the free-fermion
infrared fixed point.  For example, for the dimer operator
$F_d(x)=x/(x+1)$.  The functions $F_\U(x)$ are properties of a well-defined
renormalization group flow.  Once they have been calculated,
the behavior of the differential cross section of the
process~(\ref{process}) in the crossover region is then
\begin{equation}
  \frac{d\sigma}{d E} \sim (E_0-E)^{\Delta-\frac52}\,
  F_\U \biggl[\left(1+\frac{M_B}{M_\U} \right)
    \frac{E_0-E}{\varepsilon_0} \biggr] .
\end{equation}
There may be contributions from more than one operator $\U$ to a given
process.

It may be important to investigate the correction to the power law
coming from effects beyond the large scattering length, e.g., the
effective range or the three-body force.  This too, hopefully, could
be accomplished using techniques of effective field theory.

Nuclear reactions involving three and four neutrons in the final
states have been investigated in the searches for bound trineutron and
tetraneutrons or narrow resonances (see, e.g.,
Ref.~\cite{Kezerashvili:2016ucn,Kisamori:2016jie,Marques:2021mqf}).  Our prediction is
made under the assumption that there is no narrow resonance with
energy comparable or less than the kinetic energy of the neutrons in
the frame of their center of mass.
We have analyzed the two- and three-neutron spectra of
realistic calculations for the reactions
$^6$He$(p,p\alpha)2n$~\cite{Gobel:2021pvw},
$^3$H$(\pi^-,\gamma)3n$~\cite{Golak:2018jje},
and $^3$H$(\mu^-,\nu_\mu)3n$~\cite{Golak:2016zcw}.
These spectra show clear evidence of unnucleus behavior.
An analysis of experimental two- and four-neutron spectra 
for unnucleus behavior may become possible in the
near future~\cite{nn_scat_len_ribf_prop2018,nnnn_ribf_prop2014}.

Other types of unnuclei may be interesting to consider.  The scattering
length between two $\alpha$ nuclei is also large, so one can
consider processes where two or three $\alpha$ particles are knocked
out from a nucleus.  The unnucleus formed by three $\alpha$ particles is
where the Efimov effect takes place
\cite{Efimov:1970zz,Hammer:2010kp,Naidon:2016dpf}.
The dimension of the unnucleus
operator is complex: $\Delta = \frac52 \pm is_0$ with
$s_0\approx1.006$, so the differential cross section should have a
weak log-periodic dependence on $(E_0-E)$, crossing very near the
endpoint to $(E_0-E)^2$. However, the presence of the long-range Coulomb
repulsion complicates these systems~\cite{Higa:2008dn}.

Finally, cold atoms with fine-tuned interaction may provide another
avenue for the investigation of the universal aspects of the unnuclear
physics considered in this paper~\cite{RevModPhys.82.1225}.

\acknowledgments

The authors thank A.\ M.\ Shirokov and J.\ P.\
Vary for discussions of the multi-neutron resonances,
D.\ B.\ Kaplan for discussions and comments on an earlier
version of this manuscript,
and M.\ G\"obel and J.\ Golak for providing
data of their calculations.
H.-W.H. was supported by the Deutsche Forschungsgemeinschaft (DFG,
German Research Foundation) - Project-ID 279384907 - SFB 1245 and by the
German Federal Ministry of Education and Research (BMBF)
(Grant no.\ 05P18RDFN1).
D.T.S. was supported by the U.S.\
DOE grant No. DE-FG02-13ER41958 and a Simons Investigator grant from
the Simons Foundation.

\bibliography{few-body}

\end{document}